\documentclass[12pt]{article}
\usepackage[latin1]{inputenc}
\usepackage{cite}
\usepackage{amsmath}
\usepackage{amsfonts}
\usepackage{amssymb}
\usepackage{graphicx}
\usepackage{geometry}
\usepackage{amssymb,epsfig}
\usepackage{hyperref}
\usepackage{slashed}

\makeatletter
\renewcommand\section{\@startsection {section}{1}{\z@}%
                                 {-3.5ex \@plus -1ex \@minus -.2ex}
                                   {2.3ex \@plus.2ex}%
                                   {\normalfont\large\bfseries}}
\renewcommand\subsection{\@startsection{subsection}{2}{\z@}%
                                   {-3.25ex\@plus -1ex \@minus -.2ex}%
                                     {1.5ex \@plus .2ex}%
                                     {\normalfont\bfseries}}
\renewcommand\subsubsection{\@startsection{subsubsection}{3}{\z@}%
                                   {-3.25ex\@plus -1ex \@minus -.2ex}%
                                     {1.5ex \@plus .2ex}%
                                     {\normalfont\itshape}}
\makeatother

\def\pplogo{\vbox{\kern-\headheight\kern -29pt
\halign{##&##\hfil\cr&{\ppnumber}\cr\rule{0pt}{2.5ex}&\ppdate\cr}}}
\makeatletter
\def\ps@firstpage{\ps@empty \def\@oddhead{\hss\pplogo}%
  \let\@evenhead\@oddhead 
}
\def\maketitle{\par
 \begingroup
 \def\thefootnote{\fnsymbol{footnote}}
 \def\@makefnmark{\hbox{$^{\@thefnmark}$\hss}}
 \if@twocolumn
 \twocolumn[\@maketitle]
 \else \newpage
 \global\@topnum\z@ \@maketitle \fi\thispagestyle{firstpage}\@thanks
 \endgroup
 \setcounter{footnote}{0}
 \let\maketitle\relax
 \let\@maketitle\relax
 \gdef\@thanks{}\gdef\@author{}\gdef\@title{}\let\thanks\relax}
\makeatother

\numberwithin{equation}{section}

\newcommand{\half}{\frac{1}{2}}

\newcommand{\LL}{\mathcal{L}}

\newcommand{\del}{\partial}

\newcommand{\be}{\begin{equation}}
\newcommand{\bea}{\begin{eqnarray}}
\newcommand{\ee}{\end{equation}}
\newcommand{\eea}{\end{eqnarray}}

\newcommand{\Tr}{{\rm Tr}}

\begin{document}
 
\setcounter{page}0
\def\ppnumber{\vbox{\baselineskip14pt
}}
\def\ppdate{\footnotesize{}} \date{}

\author{Carlos Tamarit\\
[7mm]
{\normalsize  \it Perimeter Institute for Theoretical Physics, Waterloo, ON, N2L 2Y5, Canada}\\
[3mm]
{\tt \footnotesize  ctamarit at perimeterinstitute.ca }
}

\title{\bf Running couplings with a vanishing scale anomaly}
\maketitle

\begin{abstract} \normalsize
\noindent 

Running couplings can be understood as arising from the spontaneous breaking of an exact scale invariance in appropriate effective theories with no dilatation anomaly. Any ordinary quantum field theory, even if it has massive fields, can be embedded into a theory with spontaneously broken exact scale invariance in such a way that the ordinary running is recovered in the appropriate limit, as long as the potential has a flat direction.  These scale-invariant theories, however, do not necessarily solve the cosmological constant or naturalness problems, which become manifest in the need to fine-tune dimensionless parameters.
\end{abstract}

\newpage
 \section{Introduction}

Scale invariance is clearly not a property of the interactions of elementary particles in Nature. The breaking of this symmetry manifests itself not only in the presence of massive states, but also in the scaling of interaction strengths with energy, epitomized by the running of the electromagnetic \cite{GellMann:1954fq} and strong \cite{Gross:1973id,Politzer:1973fx} coupling constants, whose energy dependence has been elucidated from a plethora of experiments --see refs.~\cite{Kinoshita:1996vz} and \cite{Pich:2013sqa} for reviews.

An attractive possibility is that the breaking of scale invariance could be spontaneous, so that the Standard Model would be  a low-energy description of the broken phase of a scale-invariant theory. Indeed, scale invariance has been proposed as a possible means to solve the cosmological constant problem \cite{Antoniadis:1984kd,Wetterich:1987fm,Shaposhnikov:2008xi} as well as the Higgs hierarchy problem \cite{Bardeen:1995kv,Hempfling:1996ht,Meissner:2006zh,Chang:2007ki,Foot:2007iy,Iso:2009ss,Shaposhnikov:2008xi} --see also refs.~\cite{Iso:2012jn,Carone:2013wla,Englert:2013gz,Hambye:2013dgv} for recent analyses after the Higgs discovery \cite{Aad:2012tfa,Chatrchyan:2012ufa}-- because the scale symmetry (or possibly Weyl symmetry in theories including gravity) may force the vacuum energy to be zero and protect masses from quantum corrections.

Nevertheless, even in classically scale-invariant theories the dilatation symmetry is generically broken at the quantum level by means of the dilatation anomaly, which gives rise to a  nontrivial scaling of Green's functions with energy that can be characterized by means of the equations of the
 Renormalization Group (RG)  and the running couplings mentioned above  \cite{Callan:1970yg,Symanzik:1970rt}. This breaking of the symmetry, though successfully explaining the running, could spoil the above properties concerning naturalness and the cosmological constant, but might also potentially explain the origin of mass scales through some variant of the Coleman-Weinberg mechanism \cite{Coleman:1973jx}, triggered by the Standard Model gauge group plus additional matter \cite{Meissner:2006zh,Foot:2007iy} or a hidden gauge group which can be Abelian \cite{Hempfling:1996ht,Chang:2007ki,Iso:2009ss,Iso:2012jn,Englert:2013gz} or non-Abelian \cite{Hambye:2013dgv,Carone:2013wla}.  Approximate scale-invariance also plays an important role in models of Walking Technicolor \cite{Holdom:1984sk}; although these constructions are intrinsically nonperturbative, some of their features can be realized in a perturbative setting \cite{Grinstein:2011dq}.
 
 The scale anomaly and its important role in giving rise to the successfully measured running of couplings represent in principle serious obstacles to the task of building phenomenologically viable theories in which the breaking of scale invariance is purely spontaneous. Here, clarifying and extending previous results \cite{Shaposhnikov:2008xi,Antoniadis:1984br,Antoniadis:1984kd}, it will be argued that in the context of effective theories this is not necessarily the case, and in contrast to the usual lore that ties together running couplings and the dilatation anomaly, it is possible to have theories with exact scale invariance at the quantum level, which in the broken phase exhibit nontrivial scaling properties which in an appropriate limit reproduce the running couplings of ordinary theories with massive parameters. In this way, running couplings can be interpreted as a consequence of a spontaneously broken exact scale invariance. This preservation of the symmetry comes from using a dilatation-preserving regularization constructed by modifying the Lagrangian interactions in a way that is trivial when the regularization is removed and yet causes the cancellation of the dilatation anomaly \cite{Englert:1976ep}.  In  dimensional regularization (DR), these modifications take an extremely simple form and in the broken phase generate an infinite tower of local, higher dimensional terms --hence the characterization of these theories as effective. These higher dimensional terms have physical consequences which in principle distinguish these theories from 
those obtained using the traditional regularizations.

In theories with a scale anomaly, the lack of scale/Weyl invariance at the quantum level is understood as coming from the failure of the regulators of the quantum theory to preserve the symmetry; for example, in DR a renormalization scale $\mu$ is introduced which explicitly breaks the scale symmetry, giving rise to a scale anomaly  which generates the RG running. Nevertheless, as commented before there are regularizations that  preserve scale/Weyl invariance at the quantum level in theories involving a scalar field with a nonzero vacuum expectation value (VEV). These regularizations  were found in the context of investigations of Weyl-invariant scalar gravity \cite{Englert:1976ep,Fradkin:1978yw}. Recently, Shaposhnikov and collaborators advocated Englert et al's dilatation-preserving regularization, without demanding Weyl invariance, as a possible means to simultaneously tackle the Higgs hierarchy and cosmological constant problems \cite{Shaposhnikov:2008xi}. These theories involve a dilaton field whose vacuum expectation value generates all mass scales in the theory, and rely on DR with an unusual relation between bare and renormalized couplings which makes use of the dilaton field itself as opposed to a renormalization scale. It was argued in \cite{Shaposhnikov:2008xi} that in the limit in which the dilaton couples weakly to the other fields, one should recover the usual effective running couplings, which was supported by a one-loop calculation of the four point function in a scalar theory. Naively, this would seem puzzling, since as mentioned before 
the usual lore ties together running couplings and the scale anomaly. How can then the former arise in a theory with exact scale invariance, however spontaneously broken? Furthermore, the running coupling  constants and the scale anomaly are associated with nonzero beta functions whose knowledge permits the resummation of momentum-dependent logarithmic corrections in perturbation theory, and it is unclear whether any analogues of these can arise in a regularization devoid of an explicit renormalization scale.
 
 In the context of gravitational theories with local Weyl invariance, the idea that a Weyl-invariant theory may hide running couplings despite the lack of a scale anomaly was already put forward in refs.~\cite{Antoniadis:1984br,Antoniadis:1984kd,Tsamis:1984hh,Codello:2012sn}. Ref.~\cite{Tsamis:1984hh} proved the physical equivalence at the quantum level between Weyl-invariant theories of scalar gravity and ordinary, dilatonless theories (see also \cite{Fradkin:1978yw,Codello:2012sn}), implying indeed that the Weyl-invariant theory has hidden running couplings; nevertheless, since on the side of the scalar-gravity theory the equivalence only applies to Green's functions of Weyl-invariant operators, it cannot be readily extended to all Green's functions in scale-invariant (rather than Weyl-invariant) theories\footnote{Note also that in Weyl-invariant theories the field interactions are further restricted than in scale-invariant theories; a typical example is that the kinetic term of the dilaton is forced to be negative in Weyl-invariant equivalents of General Relativity}. Refs.~\cite{Antoniadis:1984br,Antoniadis:1984kd} tackled more directly the compatibility of running couplings with the absence of a scale anomaly in Weyl invariant theories; however, the treatment of the subject has some errors and is not very transparent  due to the introduction of an explicit renormalization scale $\mu$, which in principle would break the symmetry and generate an anomaly. This was nevertheless indirectly avoided by imposing conditions which, though wrongly stated (at the level of the effective potential, as noted by \cite{Tsamis:1984hh}, and also in regards to Ward identities, as will be seen later), are correct in spirit and would amount to $\mu$ being proportional to the VEV of the dilaton. In ref.~\cite{Codello:2012sn}, Weyl-invariant theories were regularized  by choosing dimensionful cutoffs proportional to the dilaton VEV, which again is not a transparent choice, and dilatation Ward identities were not studied \footnote{In fact, since cutoffs are unphysical the theories obtained by rewriting them as  dimensionless constants times the dilaton VEV should be physically equivalent to the theories with the usual cutoffs, which are anomalous, so that the purported scale symmetry would be unphysical. This would be similar to writing the $\mu$ parameter in DR as a constant times the dilaton VEV. See \S~\ref{sec:discussion} for related discussions.}.
 
In this paper the issue of how running couplings arise in these scale-invariant (rather than Weyl-invariant) theories with scalars, fermions and gauge fields is clarified by,  without introducing a symmetry-spoiling cutoff or renormalization scale, using dilatation Ward identities and analogues of Callan-Symanzik equations to show that, despite the absence of a scale anomaly, in the broken phase the Green's functions of these theories exhibit a nontrivial scaling behavior which, in the limit in which the  couplings of the dilaton are small but the dilaton VEV is large enough to keep the masses at finite values, gives rise to the running couplings of the corresponding dilaton-free theories with external mass scales. This extends to all orders the results of ref.~\cite{Shaposhnikov:2008xi}, which are also shown to hold for theories with gauge fields and fermions. Along the way, it is shown that the running is indeed generated by beta functions that can be calculated perturbatively in a way similar to that in theories with the usual $\mu$-dependent regularization. These beta functions are associated with the residual freedom in the regularization involving the dilaton field, and in the limit of weak couplings/large VEV of the latter coincide with the beta functions obtained for the same theory regulated with a renormalization scale $\mu$. Again, knowledge of the beta functions allows to resum perturbative corrections with logarithmic dependence on momenta. The usual logarithms of the renormalization scale appearing at finite orders of perturbation theory are traded for logarithms of the dilaton VEV times a dimensionless constant, which makes contact with the arguments of ref.~\cite{Antoniadis:1984br}.
 
 These results suggest that any theory with massive and/or massless fields can be embedded into a scale-invariant theory in which all the masses arise from a spontaneous breaking of the symmetry, in such a way that  despite the absence of a scale anomaly the spontaneous breaking ends up giving rise in the appropriate limit to the usual running couplings. However, not everything is good news. These scale-invariant theories are nonrenormalizable and might only make sense as effective theories \cite{Shaposhnikov:2009nk}. Also, paralleling the reasoning of ref.~\cite{Tsamis:1984hh} in the context of Weyl-invariant theories, it will be 
 argued that there is an intrinsic tuning in the demand for a flat direction which can be interpreted as the cosmological constant problem in disguise. Finally, it will also be pointed out that in the presence of heavy states with scale-invariant couplings, these theories are not free from unnatural tunings in the Higgs sector, since the UV parameters  have to be severely fine-tuned in order to reproduce the correct phenomenology at low energies.
 
 The discussion is organized as follows. \S~\ref{sec:ordinary_scalar} reviews how the anomalous scale Ward identities and the Callan-Symanzik equations give rise to running couplings in a simple theory with a massive scalar. In section \ref{sec:scalarplusdilaton}, the previous theory is embedded into a classically scale-invariant model by coupling the scalar field to a dilaton; the resulting scaling identities in ordinary dimensional regularization are obtained in \S~\ref{subsec:normalDR}, while their analogues in the regularization preserving scale invariance are derived in \S~\ref{subsec:phiDR}. The results are extended to theories with gauge fields and fermions in \S~\ref{sec:gauge_and_fermions}, and the nature of the dilatation symmetry is discussed in section \ref{sec:discussion}. Observations about fine-tuning are made in \S~\ref{sec:tuning}. A summary and conclusions are given in \S~\ref{sec:conclusions}.

\section{Ordinary massive scalar\label{sec:ordinary_scalar}}
 
 This section reviews how scaling Ward identities and Callan-Symanzik equations give rise to physical running couplings. For this purpose the simple case of a scalar theory with a real massive field $H$ is considered,
  \begin{align}
  {S}=\int d^4x \,\half \del_\mu H\,\del^\mu H-\frac{m^2_0}{2}H^2-\frac{\lambda_0}{4!}H^4.
  \label{eq:ordinaryscalar}
 \end{align}
 The presence of the mass parameter violates explicitly scale invariance, i.e., invariance under the transformations 
 \begin{align}
 x'&=e^{-a}x,\nonumber\\
 \label{eq:scalingtransf}H'(x')&=e^{a}H(x).
\end{align}
 If the dilatation current is defined as
 \begin{align*}
 D_\mu=\frac{\partial{\cal L}}{\partial(\del_\mu H)}\left(x^\nu\del_\nu H+H\right)-x^\mu {\cal L},
 \end{align*}
 then classically there is a nonzero divergence,
 \begin{align*}
    \partial_\mu D^\mu=\frac{\del {\cal L}}{\del H}H+2\frac{\del {\cal L}}{\del \del_\mu H}\del_\mu H-4{\cal L}=m^2 H^2.
 \end{align*}
 When regularizing the theory in dimensional regularization with space-time dimension $D=4-2\epsilon$, the mass dimensions of the field is fixed by the kinetic term to be
\begin{align*}
 [H]=\frac{D-2}{2}\equiv d,
\end{align*}
so that for consistency $\lambda_0$ acquires a nonzero mass dimension of $2\epsilon$. As usual, one may define parameters with the usual dimensions by introducing the renormalization scale $\mu$, such that 
\begin{equation}\label{eq:lambda0mu}
\lambda_0=\mu^{2\epsilon}\lambda.
\end{equation}
Modifying appropriately the dilatation transformations to reflect the new mass dimensions,
\begin{align*}
 x'&=e^{-a}x\\
 H'(x')&=e^{da}H(x),
\end{align*}
the dilatation current and its divergence equation then become
\begin{align*}\nonumber
D_\mu&=\frac{\partial{\cal L}}{\partial(\del_\mu H)}\left(x^\nu\del_\nu H+dH\right)-x^\mu {\cal L},\\
\partial_\mu D^\mu&=\frac{\del {\cal L}}{\del H}d H+\frac{\del {\cal L}}{\del \del_\mu H}(d+1)\del_\mu H-D{\cal L}=m^2 H^2+(D-4)\lambda\frac{\del {\cal L}}{\del \lambda}.
\end{align*}
Given that DR preserves the Ward identities and equations of motion at the quantum level, (i.e. DR satisfies the quantum action principle \footnote{The quantum action principle establishes the validity of identities obtained from the formal path-integral definition of the theory. If the Lagrangian ${\cal L}$ depends on a parameter $\sigma$, then the principle implies that for any operator ${\cal O}$, 
$\partial_\sigma\left\langle T{\cal O}\right\rangle=i\left\langle T{\cal O}\partial_\sigma {\cal L}\right\rangle$, where $T$ represents time ordering.} \cite{Breitenlohner:1977hr}) the previous equation holds as an operator identity inside the path integral. The contribution proportional to $4-D$ generates the usual scale anomaly, breaking the symmetry even in the massless limit \cite{Akyeampong:1973vj}. When inserted into loop diagrams, a finite contribution arises from the cancellation of the $1/\epsilon$ pole --proportional to the beta functions and anomalous dimensions-- with the $(4-D)$ factor in the anomalous scale identity. Thus an anomalous scaling behavior is associated with nonzero beta functions and anomalous dimensions, which in turn give rise to the physical running couplings \cite{Callan:1970yg,Symanzik:1970rt}. This can be seen from either the Ward identities related with the dilatation transformations or the Callan-Symanzik equations encoding the independence of physical observables from the renormalization scale $\mu$. For example, the Ward identity  corresponding to the dilatation transformations of a connected n-point function in momentum space becomes
\begin{align}\label{eq:Wardscaling1}
 \left(\sum_i p_i\frac{\del}{\del p_i}+2 m^2 \frac{\del}{\del m^2}+3n-4\right)G^{(n)}(p_i)=i\left\langle T(D-4)\lambda\partial_\lambda {\cal L}(0)\,\prod_i H(p_i)\right\rangle,
\end{align}
which may also be written, using the quantum action principle, as 
\begin{align}\label{eq:Wardscaling2}
 \left(\sum_i p_i\frac{\del}{\del p_i}+2 m^2 \frac{\del}{\del m^2}+\mu\frac{\del}{\del \mu}+3n-4\right)G^{(n)}(p_i)=0.
\end{align}
The latter equation simply ensures that, taking the $\mu$ dependence into account, the functional dependence of the Green's function $G^{(n)}$ on the dimensionful parameters is such that the correct mass dimension of $4-3n$ is achieved. The $\mu$ dependence can be evaluated with the Callan-Symanzik equations, which demand the independence of bare Green's functions with respect to the renormalization scale:
\begin{align}
\left(\mu\frac{\del}{\del \mu}+\beta_{m^2} \frac{\del}{\del m^2}+\beta_\lambda \frac{\del}{\del \lambda}+n\gamma\right)G^{(n)}(p_i)=0,
\label{eq:Callan_Symanzik}
\end{align}
with the beta and gamma functions defined as $\beta_m^2=\mu\frac{\del m^2}{\del\mu},\,\beta_\lambda=\mu\frac{\del \lambda}{\del\mu},\,\gamma=\frac{\mu}{2 Z_H}\frac{\del Z_H}{\del\mu}$, where $Z_H$ is the field renormalization constant relating the bare and renormalized field.
Combining eqs.~\eqref{eq:Wardscaling2} and \eqref{eq:Callan_Symanzik}, one gets
\begin{align}
 \left(\sum_i p_i\frac{\del}{\del p_i}+(2m^2-\beta_{m^2}) \frac{\del}{\del m^2}-\beta_\lambda \frac{\del}{\del \lambda}-n\gamma+3n-4\right)G^{(n)}(p_i)=0,
 \label{eq:Callan}
\end{align}
 As is well known, this equation implies that the Green's functions can be expressed in terms of running couplings $\bar\lambda(P),\,\bar m^2(P)$, where $P$ is a characteristic energy of the process. $\bar\lambda,\,\bar m^2$ are independent of the renormalization scale $\mu$, and they are determined by the beta functions through the following equations:
\begin{align*}
P\frac{\del \bar\lambda(P)}{\del P}&=\beta_\lambda(\bar\lambda), \quad\bar\lambda(P=\mu)=\lambda,\\
P\frac{\del \bar m^2(P)}{\del P}&=\beta_{m^2}(\bar m,\bar\lambda), \quad\bar m^2(P=\mu)=m^2.
\end{align*}
For example, if all external momenta involve a scale $P$ with $P^2\gg m^2$, the solution to the Callan-Symanzik equation for the four-point function becomes
\begin{align*}
G^{(4)}(P)=-\frac{i}{P^8}F(\bar\lambda(P))\exp\left[4\int_\mu^P d\log\frac{P'}{\mu}\gamma(\bar\lambda(P'))\right], 
\end{align*}
with $F(\lambda)=\bar\lambda+O(\bar\lambda^2)$ determined by matching with perturbation theory. Thus, the running coupling $\bar\lambda(P)$ describes the interaction strength at different energy scales.

If the theory is in a broken phase, in appropriate schemes (e.g. any mass-independent one) the counterterms are the same as in the symmetric theory \cite{Collins:1984xc}, from which it follows that the VEV $H_0$ of the field has a $\mu$-dependence given by the field renormalization factor $Z(\mu)$. This implies that the identity \eqref{eq:Callan} is replaced by
\begin{align}
 \left(\sum_i p_i\frac{\del}{\del p_i}+H_0(1+\gamma)\frac{\del}{\del H_0}+(2m^2-\beta_{m^2}) \frac{\del}{\del m^2}-\beta_\lambda \frac{\del}{\del \lambda}-n\gamma+3n-4\right)G^{(n)}(p_i)=0.
\label{eq:scalingbroken}
 \end{align}

 \section{Scalar plus dilaton\label{sec:scalarplusdilaton}}
 
  This section  studies the scaling properties of Green's functions in a theory resulting from embedding the model of eq.~\eqref{eq:ordinaryscalar} into a classically scale-invariant theory by substituting the mass parameter
 with the VEV of a field, $\phi$, which will be referred to as the dilaton:
 \begin{align}\label{eq:Sdilaton}
  {S_{inv}}=\int d^4x \,\half \del_\mu H\,\del^\mu H+\half \del_\mu \phi\,\del^\mu\phi-\frac{\lambda_0}{4!}\Big((H^2-\zeta^2\phi^2)^2+(\eta\phi)^4\Big).
 \end{align}
For $\eta=0$ one has a classical flat direction\footnote{A nonzero $\eta$ has to be considered for consistency, since the coupling is not protected by symmetries. Assuming appropriate $Z_2$ symmetries, the potential of eq.~\eqref{eq:Sdilaton} includes all renormalizable interactions. As discussed in \S~\ref{sec:tuning}, a flat direction can always be obtained by imposing the proper renormalization conditions, which to lowest order imply $\eta=0$.}, characterized by $\langle H \rangle=\zeta \langle\phi\rangle\equiv\zeta\phi_0$. The fields $H$ and $\phi$ mix yielding a massive scalar and a massless Goldstone boson. The action $ {S_{inv}}$ is classically scale invariant under the transformations of eq.~\eqref{eq:scalingtransf} supplemented by $\phi'(x')=e^a \phi(x)$. Again one may regularize the quantum theory in DR, and again a scale is needed in the relation between the bare and renormalized quartic coupling. One may introduce again a $\mu$ parameter, $\lambda_0=\mu^{2\epsilon} \lambda$, for which one gets the usual anomalous scaling with energy sourced by the dilatation anomaly; this is reviewed in \S\ref{subsec:normalDR}. On the other hand, given that the dilaton generically gets a VEV, one may use the dilation itself for the regularization \cite{Englert:1976ep}: $\lambda_0\sim (\phi)^\frac{8-2D}{D-2}\lambda$. In this case the dilatation symmetry is preserved at the quantum level, and despite the absence of an anomaly, one will still get a nontrivial scaling behavior, as reviewed in \S\ref{subsec:phiDR}.

\subsection{Usual regularization\label{subsec:normalDR}}

In this section the traditional DR regularization is considered, and the scaling behavior of Green's functions is analyzed in the broken phase in which the fields acquire VEVs, 
\begin{equation}
H=H_0+\tilde H,\phi=\phi_0+\tilde \phi,
\label{eq:VEVs}
\end{equation}
so that the classical symmetry is nonlinearly realized. In this case the equation for the divergence of the current reads
\begin{align}\label{eq:divDR} 
D_\mu&=\sum_{\varphi_i=\tilde H,\tilde\phi}\frac{\partial{\cal L}}{\partial(\del_\mu \varphi_i)}\left(x^\nu\del_\nu\varphi_i+d\varphi_i\right)-x^\mu {\cal L},\\
\nonumber\partial_\mu D^\mu&=\sum_{\varphi_i=\tilde H,\tilde\phi}\frac{\del {\cal L}}{\del \varphi_i}d\varphi_i+\frac{\del {\cal L}}{\del \del_\mu\varphi_i}(d+1)\del_\mu\varphi_i-D{\cal L}=-d\phi_0\frac{\del {\cal L}}{\del \phi_0}-dH_0\frac{\del {\cal L}}{\del H_0}+(D-4)\lambda\frac{\partial {\cal L}}{\del \lambda}.
\end{align}
As before, the scaling Ward identities pick up an anomalous piece which corresponds to a derivative with respect to the scale parameter $\mu$. For a momentum space connected Green's function with $n_H$ $H$ legs with momenta $p^H_i,1\leq i\leq n_H$, and $n_\phi$ dilaton legs with momenta $p^\phi_j, n_H+1 \leq j\leq n_H+n_\phi$, one has
\begin{align}
\label{eq:Wardscalingdilaton}
 \left(\sum_{\varphi,k} p^\varphi_k\frac{\del}{\del p^\varphi_k}+\mu\frac{\del}{\del \mu}+\sum_\varphi\left(\varphi_0\frac{\del}{\del\varphi_0}+ 3n_\varphi\right)-4\right)G^{(n_H,n_\phi)}(p^\varphi_k)=0,
\end{align}
where $\varphi=(H,\phi), \varphi_0=(H_0,\phi_0)$. The Callan-Symanzik equations for a momentum-space Green's function corresponding to $n_H$ $H$ fields and $n_\phi$ $\phi$ fields have the form 
\begin{align}
 \left(\mu\frac{\del}{\del \mu}+\beta_\lambda\frac{\del}{\del \lambda}+\beta_{\zeta}\frac{\del}{\del \zeta}+\beta_{\eta}\frac{\del}{\del \eta}+ \sum_\varphi\gamma_\varphi\left(n_\varphi- \varphi_0\frac{\del}{\del \varphi_0}\right)\right)G^{(n_H,n_\phi)}(p^\varphi_k)=0,
 \label{eq:Callan_dilaton_mu_0}
\end{align}
where again it was taken into account that, since the counterterms of the broken phase are the same as in the symmetric one, the $\mu$ dependence of the VEVs is given by the anomalous dimensions. Using the latter identity together with eq.~\eqref{eq:Wardscalingdilaton}, one gets
\begin{align}
 \left(\sum_{\varphi,k} p^\varphi_k\frac{\del}{\del p^\varphi_k}-\sum_{g=\lambda,\zeta,\eta}\beta_g\frac{\del}{\del g}
 +\sum_\varphi\left((1+\gamma_\varphi)\varphi_0\frac{\del}{\del{\varphi_0}}-\gamma_\varphi n_\varphi+3 n_\varphi\right)-4\right)G^{(n_H,n_\phi)}(p^\varphi_k)=0.
 \label{eq:scalingphi0}
\end{align}
This equation will again have solutions in terms of running couplings, the running being determined by the beta functions. In order to make contact with the scaling behavior of the theory of eq.~\eqref{eq:ordinaryscalar} in the broken phase, given by eq.~\eqref{eq:scalingbroken}, one may proceed by noticing that, given the Lagrangian of eq.~\eqref{eq:Sdilaton} with $\lambda_0$ regularized as in eq.~\eqref{eq:lambda0mu}, one has, recalling that $\phi=\phi_0+\tilde\phi$ ,
\begin{align*}
\phi_0\frac{\del}{\del \phi_0}{\cal L}=\zeta\frac{\del}{\del \zeta}{\cal L}+\eta\frac{\del}{\del \eta}{\cal L}+\tilde\phi\frac{\del}{\del \tilde\phi}V,
\end{align*}
where V is the potential of the theory. Applying the quantum action principle, this identity implies
\begin{align}
\label{eq:QAderivatives}
\phi_0\frac{\del}{\del \phi_0} =\zeta\frac{\del}{\del\zeta}+\eta\frac{\del}{\del\eta} +i\left(\tilde\phi\frac{\del{V}}{\del \tilde\phi}\right)(0)\ast,
\end{align}
where $\left(\tilde\phi\frac{\del{V}}{\del \tilde\phi}\right)(0)\ast$ represents an insertion of the corresponding operator into the appropriate Green's function, e.g.
\begin{align*}
 \left(\tilde\phi\frac{\del{V}}{\del \tilde\phi}\right)(0)\ast G^{n_H,n_\phi}(p^\varphi)=\left\langle T\left(\tilde\phi\frac{\del{V}}{\del \tilde\phi}\right)(0)\prod_{i} \prod_{j} \tilde H(p^H_i)\tilde\phi(p^\phi_j)\right\rangle.
\end{align*}
The scaling Ward identities \eqref{eq:scalingphi0}, together with eq. \eqref{eq:QAderivatives} --being careful to restore the proper $D$-dependence coming from the $D$-dimensional Ward identity \eqref{eq:divDR} in the contributions involving insertions of composite operators, whose renormalization is subtle-- yield
\begin{align}
\nonumber & \left(\sum_{\varphi,k} p^\varphi_k\frac{\del}{\del p^\varphi_k}\!+\!H_0\frac{\del}{\del H_0}+\sum_{\tilde g=\zeta,\eta}(\tilde g\!-\!\beta_{\tilde g}) \frac{\del}{\del \tilde g}\!-\beta_\lambda \frac{\del}{\del \lambda}
+\sum_\varphi\left(\gamma_\varphi\varphi_0\frac{\del}{\del{\varphi_0}}-\gamma_\varphi n_\varphi+3n_\varphi\right)-4\right)\times\\
&\times G^{(n_H,n_\phi)}(p^\varphi_k)=-i\frac{D-2}{2}\left(\tilde\phi\frac{\del{V}}{\del \tilde\phi}\right)(0)\ast G^{n_H,n_\phi}(p^\varphi_k).
  \label{eq:Callan_dilaton_mu}
\end{align}
The right hand side, at least before implementing renormalization, equals $\frac{D-2}{2}$ times a sum obtained from the usual diagrams as follows: for each of these diagrams, one adds up as many contributions as possibilities of singling out one vertex --internal or external-- in the parent diagram,  each contribution being equal to the latter times the number of dilaton fields in the singled out vertex. Thus, only diagrams with internal or external dilaton legs can contribute.
One may consider the limit in which the dilaton decouples while giving an appreciable mass to the $H$ field. In this limit one has $\zeta,\eta\ll 1$, while $\zeta \phi_0$ is kept finite in order to generate the $H$ mass.  $\eta\phi_0$ remains negligible, as follows from requiring the existence of a flat direction. In this limit,  diagrams with internal or external dilaton legs can be ignored, so that $\gamma_\phi,\eta,\beta_\eta$  and the right hand side  can be neglected. One may define a mass scale for $H$ as $m^2=\frac{1}{6}\lambda\zeta^2\phi_0^2$; taking into account the fact that if the dilaton decouples the dependence of the Green's functions on $\zeta$ only comes from factors of $\lambda\zeta^2\phi_0^2$, then one may recast the dependence of the Green's functions on $\zeta,\lambda$ into a dependence on $m^2,\lambda$, so that
\begin{align*}
& \zeta\frac{\del}{\del \zeta}\rightarrow2m^2\frac{\del}{\del m^2},\\
 &\beta_\zeta\frac{\del}{\del\zeta}+\beta_\lambda\frac{\del}{\del\lambda}\rightarrow \beta_{m^2}\frac{\del}{\del m^2}+\beta_\lambda\frac{\del}{\del\lambda},\quad \beta_{m^2}=\frac{1}{6}\beta_\lambda\zeta^2\phi_0^2+\frac{1}{3}\beta_\zeta \zeta \lambda \phi_0^2= \mu\frac{\del m^2}{\del \mu}+O(\gamma_\phi).
\end{align*}
Hence in this limit one may write for Green's functions with only external $H$ fields
\begin{align*}
  &\!\left(\sum_i p_i\frac{\del}{\del p_i}\!+\!(1\!+\!\gamma_H)H_0\frac{\del}{\del H_0}\!+\!(2m^2\!-\!\beta_{m^2}) \frac{\del}{\del m^2}\!-\!\beta_\lambda \frac{\del}{\del \lambda}\!-\!n_H\gamma_H\!+\!3n_H\!-\!4\!\right)G^{(n_H)}(\lambda,m,;p_i)\sim \\
  &\sim0,\quad \zeta\ll1, \,\zeta\phi_0 \text{ finite},
\end{align*}
which reproduces eq.~\eqref{eq:scalingbroken} in the theory without a dilaton.

\subsection{Symmetry-preserving regularization\label{subsec:phiDR}}

One may also regularize the theory in a way that preserves the scale invariance;  instead of using the renormalization scale parameter $\mu$, one may use the dilaton field itself \cite{Englert:1976ep,Shaposhnikov:2008xi},
\begin{equation}
 \lambda_0=(\xi^\frac{D-2}{2}\phi)^{\frac{8-2D}{D-2}}\lambda,
 \label{eq:lambda0phi}
\end{equation}
with $\xi$ a dimensionless constant \footnote{The exponent of $\xi$ inside the parenthesis is chosen in order to obtain a total power of $\xi^{2\epsilon}$ similar to the power of $\mu^{2\epsilon}$ in the usual regularization. As a consequence of this, the beta functions and anomalous dimensions appearing in both regularizations come exclusively from the $1/\epsilon$ divergences of the diagrams after proper subtractions, which makes direct comparisons easier. Different powers of $\xi$ would define different renormalization schemes.}. In this case, from the form of the dilatation current in the first line of eq.~\eqref{eq:divDR} it immediately follows that in an unbroken phase the dilatation current is conserved at the quantum level, $D_\mu J^\mu=0$. This equation in DR is an operator identity, and will hold in the renormalized theory for a renormalization scheme such as minimal subtraction that preserves the Ward identities and equations of motion, so that no diagrams need to be computed. Things are less trivial if one considers the invariance under conformal transformations and considers the trace of the energy momentum tensor, as in ref.~\cite{Armillis:2013wya}, which presents calculations supporting conformal invariance (see also ref.~\cite{Gretsch:2013ooa}). This would be in line with the recent proofs linking unbroken scale and conformal invariance in unitary quantum field theories \cite{Komargodski:2011xv,Luty:2012ww,Fortin:2012cq}; the theories dealt with here could be a novelty since they are only well defined perturbatively in the broken phase.

Indeed, the regularization using the dilaton field is only practical when the dilaton gets a VEV and one may expand the quantum fluctuations of $\phi$ around it in order to obtain a local expansion of the Lagrangian. In this case the dilatation symmetry is again nonlinearly realized, but there is still no anomalous contribution,
\begin{align*}
 \del_\mu D^\mu=-d\phi_0\frac{\del {\cal L}}{\del \phi_0}-dH_0\frac{\del {\cal L}}{\del H_0}.
\end{align*}
This implies Ward identities for connected, renormalized Green's functions of the form (using the same notation as in eq.~\eqref{eq:Wardscalingdilaton})
\begin{align}
\label{eq:Wardscaling}
 \left(\sum_{\varphi,k} p^\varphi_k\frac{\del}{\del p^\varphi_k}+\sum_\varphi\left(\varphi_0\frac{\del}{\del\varphi_0}+3 n_\varphi\right)-4\right)G^{(n_H,n_\phi)}(p^\varphi_k)=0,
\end{align}
which just convey the fact that $G^{(n)}$ has a functional dependence on $p_i$ and $\phi_0$ ensuring that the proper mass dimension is achieved; however, there is no ``anomalous'' $\mu$ dependence in this case. It should be noted that eq.~\eqref{eq:Wardscaling}, in comparison with the Ward identity appearing in ref.~\cite{Antoniadis:1984br} (eq.~(4.16) therein), apart from the difference in mass dimensions due to the fact that eq.~\eqref{eq:Wardscaling} deals with connected rather than one-particle-irreducible Green's functions, involves $\varphi_0\frac{\del}{\del\varphi_0}G^{(n_H,n_\phi)}$ rather than a term  proportional to a Green's function with a $\phi_0$ insertion. The expression of ref.~\cite{Antoniadis:1984br} and some the ones derived from it are only valid to lowest order of insertions of $\phi_0$, which is nevertheless the regime in which explicit computations were performed in that reference. 

The scaling Ward identities by themselves make no prediction about  physical running couplings, although they certainly allow for $\log \frac{p}{\phi_0}$ corrections for dimensionless couplings. Additional input may be obtained from considering the ambiguity in the $\xi$ parameter in the relation~\eqref{eq:lambda0phi}. Note that, as happens with the $\mu$ dependence in the usual regularization, the dependence on $\xi$ vanishes in four dimensions and should be unphysical. By demanding independence of the bare Green's functions with respect to $\xi$ --or equivalently, requiring that $\xi$ does not affect physical results, so that the renormalized Green's functions only depend on it through unphysical field renormalization factors-- one gets an analogue of the usual Callan-Symanzik equations for the renormalized Green's functions, despite the absence of $\mu$:
\begin{align}
 \left(\xi\frac{\del}{\del \xi}+\hat\beta_\lambda\frac{\del}{\del \lambda}+\hat\beta_{\zeta}\frac{\del}{\del \zeta}+\hat\beta_{\eta}\frac{\del}{\del \eta}+\sum_\varphi\hat\gamma_\varphi\left(n_\varphi-\varphi_0\frac{\del}{\del \varphi_0}\right)+n.r.\right)G^{(n_H,n_\phi)}(p^\varphi_k)=0.
 \label{eq:Callandilaton}
\end{align}
In the previous formula, $\hat\beta_{\zeta}=\xi\frac{\del\zeta}{\del\xi},\,\hat\beta_{\eta}=\xi\frac{\del\eta}{\del\xi},\,\hat\beta_\lambda=\xi\frac{\del \lambda}{\del\xi},\,\hat\gamma_{H/\phi}=\frac{1}{2Z_{H/\phi}}\xi\frac{\del Z_{H/\phi}}{\del\xi}$. The contribution ``n.r'' represents  terms that would come from the beta functions of the couplings of additional operators that may have to be added to the theory for consistency if it is nonrenormalizable (as indeed seems to be the case \cite{Shaposhnikov:2009nk}).

Now, given the Lagrangian in eq.~\eqref{eq:Sdilaton}, with $\phi=\phi_0+\tilde\phi$, $\phi_0$ being constant, taking into account eq.~\eqref{eq:lambda0phi} and using the quantum action principle in DR, it follows that, similarly to eq.~\eqref{eq:QAderivatives},
\begin{align}
 \phi_0\frac{\del}{\del \phi_0}=\frac{2}{D-2}\,\xi\frac{\del}{\del \xi}+\zeta\frac{\del}{\del \zeta}+\eta\frac{\del}{\del \eta}+i\left(\tilde\phi\frac{\del{V}}{\del \tilde\phi}\right)(0)\ast.
 \label{eq:partialx0}
\end{align}
Combining eqs.~\eqref{eq:Wardscaling}, \eqref{eq:Callandilaton}
and \eqref{eq:partialx0}, reinstating the correct $D$-dependence in the contributions involving operator insertions, one obtains for renormalized Green's functions that
\begin{align}\nonumber
  &\left(\sum_{\varphi,k} p^\varphi_k\frac{\del}{\del p^\varphi_k}\!+\!H_0\frac{\del}{\del H_0}\!+\sum_{\tilde g=\zeta,\eta}\!(\tilde g\!-\!\hat\beta_{\tilde g}) \frac{\del}{\del \tilde g}\!-\!\hat\beta_\lambda \frac{\del}{\del \lambda}
\!+\!\sum_\varphi\left(\hat\gamma_\varphi\varphi_0\frac{\del}{\del{\varphi_0}}\!-\!\hat\gamma_\varphi n_\varphi+3n_\varphi\right)\!-\!4\!+\!n.r.\!\right)\!\!\times\\
&\times G^{(n_\varphi)}(p^\varphi_k)= -i\frac{D-2}{2}\left(\tilde\phi\frac{\del{V}}{\del \tilde\phi}\right)(0)\ast G^{n_\varphi}(p^\varphi_k) .
  \label{eq:Callan_dilaton_phi}
\end{align}
This equation, predicting a nontrivial scaling with momentum despite the absence of anomaly, looks the same as eq.~\eqref{eq:Callan_dilaton_mu}, obtained using the traditional regularization, but with the beta functions and anomalous dimensions substituted by their hatted counterparts. Also, it should be remembered that the potential appearing on the r.h.s. depends on the choice of regularization, so that in the scale-invariant case it includes higher dimensional interactions.

The functions $\beta,\gamma$ in eq.~\eqref{eq:Callan_dilaton_mu} and their counterparts $\hat\beta,\hat\gamma$ in eq.~\eqref{eq:Callan_dilaton_phi} are calculated by demanding independence of the bare Green's functions with respect to $\mu$ and $\xi$, respectively, as in eqs.~\eqref{eq:Callan_dilaton_mu_0} and \eqref{eq:Callandilaton}. These beta functions are fixed by the divergences of the diagrams and the choice of subtraction scheme. With the relations between bare and renormalized quartic coupling in eqs.~\eqref{eq:lambda0mu} and \eqref{eq:lambda0phi}, it can be seen that for both regularizations the beta functions only depend on the $1/\epsilon$ poles of diagrams after the subtraction of lower order counterterms. Now, since the Lagrangians of the theories differ by $O(\epsilon)$ higher-dimensional interactions, their one-loop divergences are identical. Starting at two-loop order, the evanescent (vanishing in $D=4$ dimensions) interactions in the theory preserving scale invariance might give rise to new $1/\epsilon$ and higher divergences, which will either modify the beta functions of the couplings of the renormalizable operators or correspond to  higher dimensional local operators which will get their own beta functions.
 
Therefore the functions $\beta,\gamma$ and $\hat\beta,\hat\gamma$ are expected to differ starting at two-loops, so that the scaling with momenta of the Green's functions is different in both cases. However, the difference is controlled by dilaton interactions, which are suppressed either by $\zeta,\eta$ or inverse powers of $\phi_0$. In fact, in the limit $\zeta,\eta\ll1$, $\zeta\phi_0$ finite and $\eta\phi_0$ negligible, in which the interactions of the dilaton with $H$ can be neglected, the functions  $\hat\beta,\hat\gamma$ will coincide with their unhatted counterparts. As a consequence, repeating the same arguments that were used after eq.~\eqref{eq:Callan_dilaton_mu}, one concludes that in this limit one will again reproduce the scaling behavior of eq.~\eqref{eq:Callan}, corresponding to the theory  with massive parameters and no dilaton and giving rise to the usual running couplings.

The appearance of the usual logarithmic momentum-dependence in the four point Green's function with $H$ fields was seen explicitly at one-loop in ref.~\cite{Shaposhnikov:2008xi}, yielding the usual effective $\lambda(p)$. In that calculation the parameter $\xi$ appearing in the regularization was kept fixed and, in the absence of a renormalization scale $\mu$, it was unclear whether one could associate $\lambda(p)$ with a beta function with which to effectively resum logarithmic corrections in perturbation theory.  This is clarified by the scaling identities in this section, which show how the usual physical running arises from the Ward identities and the analogues of the Callan-Symanzik equations, generalizing the result of ref.~\cite{Shaposhnikov:2008xi} to higher orders and arbitrary Green's functions, and showing that indeed one may define beta functions that allow to resum logarithmic corrections in the usual way.

One can therefore conclude that in a theory with exact scale invariance achieved through interactions with a scalar dilaton, the spontaneous breaking of the symmetry still leads to a nontrivial scaling of Green's functions with energy, which in the limit of weak couplings of the dilaton field reproduces the running coupling constants of dilatonless theories with mass parameters.

\section{Theories with gauge fields and fermions\label{sec:gauge_and_fermions}}

The results of the previous section show that one may embed a scalar theory with mass parameters and anomalous scaling behavior into a theory that is scale invariant at the quantum level, so that the mass parameters arise from a spontaneous breaking of the symmetry, which in turn gives rise to a nontrivial scaling behavior of the Green's functions, which reproduces the usual running in an appropriate limit. Though this was explicitly derived for a theory with just one scalar field apart from the dilaton, similar arguments will apply for theories with more general interactions and field content. The only important restriction is that a nonzero VEV for $\phi$ is required for perturbation theory to be meaningful, so that the potential must have a flat direction.

This construction of scale-invariant theories can be generalized to include gauge fields and fermions \cite{Armillis:2013wya}, again by introducing a dilaton with a flat direction, regularizing the theory in DR and using the dilaton itself in the relation between bare and renormalized couplings. For a bare gauge coupling $g_0$ and a bare Yukawa coupling $y_0$, this implies 
\begin{equation}\label{eq:lambda0gaugefer}
g_0=(\xi^\frac{D-2}{2}\phi)^{\frac{8-2D}{D-2}} g, \quad y_0=(\xi^\frac{D-2}{2}\phi)^{\frac{4-D}{D-2}} y.
\end{equation}
One may consider for example a theory of nonabelian gauge fields coupled to fermions, which in turn are coupled to dilatons, as follows,
\begin{align*}
 {\cal L}&=\frac{1}{4}\Tr F_{\mu\nu}F^{\mu\nu}+\frac{1}{2}\partial_\mu\phi\partial^\mu\phi+\bar\psi i\slashed{D}\psi-y_0\phi \bar\psi\psi-\frac{1}{4!}(\eta\phi)^4,\\
 F_{\mu\nu}&=\partial_\mu A_\nu-\partial_\nu A_\mu-ig_0[A_\mu,A_\nu], \quad D_\mu=\partial_\mu-ig_0A_\mu.
\end{align*}
Again, one can guarantee a flat direction for $\phi$ (choosing $\eta=0$ to lowest order), and the dilaton's VEV generates a mass for the fermions. Regularizing the theory in a scale-invariant way using eqs.~\eqref{eq:lambda0gaugefer}, one gets Ward identities for connected, renormalized Green's functions of the form
\begin{align}
\label{eq:Wardscaling3}
 \left(\sum_{\varphi,k} p^\varphi_k\frac{\del}{\del p^\varphi_k}+\phi_0\frac{\del}{\del\phi_0}+\sum_\varphi\left(c_\varphi n_\varphi\right)-4\right)G^{(n_\varphi)}(p^\varphi_k)=0,
 \end{align}
where $\varphi=(A_\mu,\psi,\bar\psi,\phi), \,n_\varphi=(n_A,n_\psi,n_{\bar\psi},n_\phi),\,c_\phi=(3,5/2,5/2,3)$ and $p^\varphi_k,\,k=1,\dots,n_\varphi$ are the momenta of the $k$-th leg involving the field $\varphi$. Demanding again the independence of physical results from $\xi$ by requiring all $\xi$ dependence to appear in the field renormalization factors, one gets
\begin{align}
\label{eq:Callandilatonfergauge}
 \left(\xi\frac{\del}{\del \xi}+\hat\beta_g\frac{\del}{\del g}+\hat\beta_{y}\frac{\del}{\del y}+\hat\beta_\eta\frac{\del}{\del \eta}+\sum_\varphi\hat\gamma_\varphi n_\varphi-\hat\gamma_{\phi}\phi_0\frac{\del}{\del \phi_0}+n.r.\right)G^{(n_H,n_\phi)}(p^\varphi_k)=0,
\end{align}
with $\hat\beta_{g}=\xi\frac{\del g}{\del\xi},\,\hat\beta_y=\xi\frac{\del y}{\del\xi},\,\hat\beta_{\eta}=\xi\frac{\del \eta}{\del\xi},\,\hat\gamma_{\varphi}=\frac{1}{2 Z_{\varphi}}\xi\frac{\del Z_{\varphi}}{\del\xi}$. Combining the above identities and using the analogue of eq.~\eqref{eq:partialx0},
\begin{align*}
 \phi_0\frac{\del}{\del \phi_0}=\frac{2}{D-2}\xi\frac{\del}{\del \xi}+y\frac{\del}{\del y}+\eta\frac{\del}{\del \eta}+i\left(\tilde\phi\frac{\del{V}}{\del \tilde\phi}\right)(0)\ast,
\end{align*}
the following identity is obtained:
\begin{align*}\nonumber
  &\left(\sum_{\varphi,k} p^\varphi_k\frac{\del}{\del p^\varphi_k}\!+\!\!\sum_{\tilde g=y,\eta}\!(\tilde g\!-\!\hat\beta_{\tilde g}) \frac{\del}{\del \tilde g}\!-\!\hat\beta_g \frac{\del}{\del g}
\!+\!\hat\gamma_{\phi}\phi_0\frac{\del}{\del \phi_0}\!+\!\sum_\varphi n_\varphi\left(c_\varphi\!-\!\hat\gamma_\varphi\right)\!-\!4\!+\!n.r.\right)G^{(n_\varphi)}(p^\varphi_k)=\\
  &= -i\frac{D-2}{2}\left(\tilde\phi\frac{\del{V}}{\del \tilde\phi}\right)(0)\ast G^{(n_\varphi)}(p^\varphi_k) .
\end{align*}
By reasoning as in the previous sections, one can see that in the limit $y,\eta\ll1$, keeping $y\phi_0$ finite and $\eta\phi_0$ negligible (as needed if there is a flat direction), then $\eta,\beta_\eta,\gamma_\phi$ and the right-hand side can be neglected and one recovers the scaling behavior of the corresponding dilatonless theory with gauge fields and massive fermions with $m_f=y\phi_0$ and the usual beta functions:
\begin{align*}\nonumber
  &\left(\sum_{\varphi,k} p^\varphi_k\frac{\del}{\del p^\varphi_k}+(m_f-\beta_{m_f}) \frac{\del}{\del m_f}-\beta_g \frac{\del}{\del g}
+\sum_\varphi n_\varphi\left(c_\phi -\gamma_\varphi\right)-4\right)G^{(n_\varphi)}(p^\varphi_k)=0,
\end{align*}
The same conclusions will apply in a theory containing gauge, fields, fermions and scalars altogether.


\section{Is the dilatation symmetry physical?\label{sec:discussion}}

The results presented here contradict the common knowledge about running coupling constants, which attributes their nontrivial  behavior under energy rescalings to the breaking of dilatation symmetry at the quantum level. Why is this no longer the case? In fact, are theories that preserve scale invariance physically equivalent to the ones defined with the usual $\mu$-dependent regularization? 

The running with energy in the usual case can be traced back to the scale anomaly because the anomalous piece in the dilatation Ward identity is equivalent to a $\mu$-derivative --see eqs.~\eqref{eq:Wardscaling1} and \eqref{eq:Wardscaling2}-- which through the Callan-Symanzik equations is then associated with the beta functions and anomalous dimensions. In the case with exact scale invariance, there is no anomalous piece in the scaling Ward identities, as is clear from eqs.~\eqref{eq:Wardscaling}, \eqref{eq:Wardscaling3}. The nontrivial scaling arises from using the quantum action principle to trade the dependence on the dilaton VEV $\phi_0$ with the dependence on the dilaton fluctuations and couplings as well as on the parameter $\xi$ appearing in the regularization (eqs.~\eqref{eq:lambda0phi} and \eqref{eq:lambda0gaugefer}). It is this $\xi$ dependence which, through the analogues of the Callan-Symanzik equations \eqref{eq:Callandilaton} and \eqref{eq:Callandilatonfergauge}, brings about the beta functions and anomalous dimensions that will define the physical running with the energy scale. Therefore the running appears because part of the $\phi_0$ dependence lies in the evanescent powers of the dilaton introduced by the regularization. Another way to see this relies on rewriting the dependence on $\phi_0$ as
\begin{align*}
\frac{\del}{\del \phi_0}\equiv\frac{\del}{\del \phi_0^{clas}}+\frac{\del}{\del \phi_0^{reg}},
\end{align*}
where $\frac{\del}{\del \phi_0^{clas}}$ traces the contribution of $\phi_0$ to tree-level masses and dimensionful couplings in $D=4$, while $\frac{\del}{\del \phi_0^{reg}}$ follows the dependence on $\phi$ coming from the regularization (see eqs.~\eqref{eq:lambda0phi}, \eqref{eq:lambda0gaugefer}). The quantum action principle implies, taking the scalar theory of \S~\ref{subsec:phiDR} as an example and recalling eq.~\eqref{eq:VEVs},
\begin{align}\label{eq:phi0regQAP}
 \phi_0^{reg}\frac{\del}{\del \phi_0^{reg}}=i\left(\frac{8-2D}{D-2}\lambda\partial_\lambda\LL+\tilde\phi^{reg}\frac{\delta}{\delta\tilde\phi^{reg}}V\right)(0)\ast,
\end{align}
so that the Ward identity \eqref{eq:Wardscaling} becomes
\begin{align}
\nonumber& \left(\sum_{\varphi,k} p^\varphi_k\frac{\del}{\del p^\varphi_k}+H_0\frac{\del}{\del H_0}+\phi_0^{clas}\frac{\del}{\del\phi_0^{clas}}+
 3\sum_\varphi n_\varphi-4\right)G^{(n_H,n_\phi)}(p^\varphi_k)=\\
 \label{eq:scalingphi0phys}&=i\left((D-4)\lambda\partial_\lambda\LL-\frac{D-2}{2}\tilde\phi^{reg}\frac{\del}{\del\tilde\phi^{reg}}{V}\right)(0)\ast G^{(n_H,n_\phi)}(p^\varphi_k).
\end{align}
The first contribution in the second line, which comes from the $\phi_0$ dependence of the regulator (see eq.~\eqref{eq:phi0regQAP}), looks exactly like the anomalous piece in the theory regulated in the usual way (see for example eq.~\eqref{eq:Wardscaling1}), and in fact, being equivalent to $-\xi\frac{\del}{\del\xi}$, generates the beta and gamma function terms in the following identity obtained by combining eq.~\eqref{eq:scalingphi0phys} with eq.~\eqref{eq:Callandilaton}:
\begin{align}\nonumber
  &\left(\sum_{\varphi,k} p^\varphi_k\frac{\del}{\del p^\varphi_k}-\!\sum_{g=\lambda,\zeta,\eta}\!\hat\beta_g \frac{\del}{\del g}
\!+\!\sum_\varphi\!\left(\varphi_0^{clas}\frac{\del}{\del \varphi_0^{clas}}\!+\!\hat\gamma_\varphi\varphi_0\frac{\del}{\del{\varphi_0}}\!-\! \hat\gamma_\varphi n_\varphi\!+\!3 n_\varphi\!\right)\!\!-\!4\!+\!n.r.\!\right)G^{(n_\varphi)}(p^\varphi_k)\\
  &=-i\frac{D-2}{2}\tilde\phi^{reg}\frac{\del}{\del\tilde\phi^{reg}}{V}(0)\ast G^{(n_\varphi)}(p^\varphi_k).
  \label{eq:anomalous_pys_dilat}
\end{align}
The r.h.s. in the previous equation is given before renormalization by $\frac{D-2}{2}$ times a sum of diagrams involving higher dimensional vertices coming from the expansion of the evanescent powers of the dilaton field, each term in the sum coming from singling out one of these vertices and multiplying the diagram by the number of dilaton legs in the chosen vertex.
From  eq.~\eqref{eq:anomalous_pys_dilat}  it is apparent that, if one forces a distinction between the $\phi_0$ dependence coming from classical masses and couplings and that coming from evanescent powers in the Lagrangian, then from the point of view of the ``classical'' scales $\phi_0^{clas}$ the dilatation symmetry is anomalous, since for a true ``classical'' dilatation symmetry one should have
\begin{align}\label{eq:physicaldilatation}
 \left(\sum_{\varphi,k} p^\varphi_k\frac{\del}{\del p^\varphi_k}+\sum_\varphi\left(\varphi_0^{clas}\frac{\del}{\del\varphi_0^{clas}}+3 n_\varphi\right)-4+n.r.\right)G^{(n_\varphi)}(p^\varphi_k)=0.
\end{align}
If one could argue that $\phi_0^{clas}$ is the only true physical scale, this could be interpreted as a hint that the dilatation symmetry using the full $\phi_0$ dependence (which implies eq.~\eqref{eq:Wardscaling} as opposed to \eqref{eq:physicaldilatation}) is  not a dilatation in the usual sense, and the running comes from the fact that the regulator itself, being dependent on the dilaton, is not invariant under rescalings.  
However, the argument fails because the separation between $\phi_0^{clas}$ and $\phi_0^{reg}$ is unphysical itself,  since it is based on the classical Lagrangian rather than on the full quantum-effective action, for which evanescent powers of $\phi_0^{reg}$ in the tree-level interactions can yield finite, physical contributions after renormalization. Calling $\phi_0^{reg}$ unphysical would be justified if the theories regularized in a scale-invariant way were physically equivalent to those defined with the usual $\mu$-dependent regularization. This would have been the case had we written, for example,
\begin{align}
\label{eq:lambdazetaphi}
 \lambda_0=(\zeta \phi_0)^{2\epsilon}
\end{align}
instead of eq.~\eqref{eq:lambda0phi}. In this case we would have gotten Ward identities appearing to preserve spontaneously broken scale invariance, i.e. of the form of eq.~\eqref{eq:Wardscaling}. This is misleading since there would be no formal dilatation symmetry away from the unbroken phase.  In fact, when distinguishing as above between $\phi_0^{clas}$ and $\phi_0^{reg}$, then it is easily seen that $\phi_0^{reg}$ is truly unphysical in this case. This is so because the dependence on the latter does not arise from expanding a dynamical field as in eq.~\eqref{eq:VEVs}, so that $\phi_0^{reg}\frac{\partial}{\partial\phi_0^{reg}}$ can be traded with $\zeta\frac{\partial}{\partial \zeta}$ --in contrast with eq.~\eqref{eq:phi0regQAP}-- and $\zeta$ is an unphysical parameter. Actually, $\zeta\frac{\partial}{\partial \zeta}$ ends up generating the ordinary anomalous scaling. Thus the purported scale invariance would be unphysical, which is not surprising since eq.~\eqref{eq:lambdazetaphi} is a simple rewriting of the usual $\mu$-regularization of eq.~\eqref{eq:lambda0mu} by considering $\mu\propto\phi_0$, and physics is independent of the choice of renormalization scale. Similar arguments cast doubts about the nature of the claimed Weyl invariance in the theories of ref.~\cite{Codello:2012sn}, in which the cutoffs are taken to be proportional to the dilaton VEV.

In contrast with the previous situation, in the dilatation-preserving regularization arising from  eq.~\eqref{eq:lambda0phi} all dependence in $\phi_0$ comes from expanding a dynamical dilaton field as in eq.~\eqref{eq:VEVs} --so that at least formally there is still a dilatation symmetry in the unbroken phase-- and then not all of the dependence on $\phi_0^{reg}$ is unphysical. In particular, the dilaton dependence in eq.~\eqref{eq:lambda0phi} gives rise to new, evanescent higher dimensional interactions which will modify the finite parts of Green's functions at one loop and beyond, and will also change the divergent parts from two loops onwards, consequently affecting the scaling behavior through contributions to the beta functions. This makes these theories  physically inequivalent to those using the traditional regularization, making the dilatation symmetry physically relevant.  A clear manifestation of this is the form of the effective potential. Consider for example the case in which the dilaton is the only scalar field, interacting with fermions at the classical level. In the usual regularization, the breaking of scale invariance at the quantum level gives rise to $\mu$-dependent radiative corrections which produce an instability, as an example of dimensional transmutation gone wrong. The minimization conditions of the potential
yield  a relation between dimensionless and dimensionful parameters of the form \cite{Coleman:1973jx}
\begin{align}
\label{eq:transmutation}
 f\left(g_i,\frac{\langle\phi\rangle}{\mu}\right)=0,
\end{align}
where $g_i$ represents the dimensionless couplings. This equation allows to trade one dimensionless coupling for the dimensionful field VEV. A frequent choice of renormalization scale is $\mu=\phi$ \cite{Coleman:1973jx}, which makes the $\mu$-dependence disappear from eq.~\eqref{eq:transmutation}, leading to an apparent fine-tuning condition for the dimensionless couplings which is however an artifact of the choice of renormalization scale, hiding the VEV dependence. Nevertheless, in this case the equation corresponding to eq.~\eqref{eq:transmutation} only has solutions for $\lambda<0$, so that the critical point is unstable.  Things are different when using the regularization that preserves true scale invariance, in which case the symmetry demands the renormalized potential to simply be of the form
\begin{align}
\label{eq:Vdilaton}
 V=\tilde f (g_i) \phi^4
\end{align}
for some function $\tilde f$ of the dimensionless couplings. This forbids the appearance of terms depending logarithmically on $\phi_0$, which would play the role analogous to $\mu$ in eq.~\eqref{eq:transmutation}\footnote{Interestingly, the logarithms depending on $\phi_0$ will cancel between the finite parts of the diagrams and the contributions coming from the counterterms, which in order to preserve scale-invariance in D dimensions  involve evanescent powers of the dilaton.}. Thus, no logarithms appear and dimensional transmutation does not take place in the usual way --which is a good thing, in this case, since no instabilities appear. In fact, generically the only possible vacuum is $\phi=0$, unless one forces $\tilde f (g_i)=0$, in which case there is a flat direction. Note that in this case the identity $\tilde f (g_i)=0$ is a true fine-tuning condition between dimensionless couplings, completely independent of any dimensionful scale, as opposed to the identities derived in the usual theories with dimensional-transmutation when setting $\mu=\langle\phi\rangle$ as explained above. This tuning can be interpreted as the cosmological constant problem in disguise \cite{Tsamis:1984hh}, as will be elaborated further in the next section.

The separation between $\phi_0^{clas}$ and $\phi^{reg}$ above allows to make comparisons with refs.~\cite{Antoniadis:1984kd,Antoniadis:1984br}, which also tackled the compatibility of running with a zero scale anomaly. These papers only dealt with ``supraconformal'' dilaton interactions, i.e. such that the dilaton dependence vanishes in $D=4$, for which $\phi_0^{clas}=0$. Unlike in the formalism of this article, an explicit renormalization scale was introduced by writing evanescent powers of $\phi$ as in eq.~\eqref{eq:lambda0phi} in the following manner:
\begin{align}
\label{eq:muphi0}
\xi^\frac{4-D}{2}(\phi_0^{reg}+\tilde\phi^{reg})^\frac{4-D}{D-2}\rightarrow
\mu^\frac{4-D}{D-2}\left(1+\frac{\sigma}{v_0}\right)^\frac{4-D}{D-2}.
\end{align}
The authors proceeded by demanding compatibility of the dilatation Ward identity --eq.~(4.16) in \cite{Antoniadis:1984br}, which contains errors, see comments around eq.~\eqref{eq:Wardscaling} of this article-- with the restrictions imposed upon Green's functions by demanding the correct mass dimension (eq.~(4.17) in \cite{Antoniadis:1984kd}). In doing so, they arrived at
eq.~(4.18) in ref.~\cite{Antoniadis:1984br}, also appearing as eq.~(8) in ref.~\cite{Antoniadis:1984kd}. This condition is incorrect, and doing things properly one would arrive at the need to make the following derivative operators equivalent when acting on Green's functions:
\begin{align*}
 \mu\frac{\partial}{\del \mu}+v_0\frac{\partial}{\del v_0}\leftrightarrow v_0\frac{\partial}{\del v_0}.
\end{align*}
For nonzero beta functions (i.e. nonzero contributions of  $\mu\frac{\partial}{\del \mu}$) , the former can be satisfied by considering $\mu$ not to be independent of $v_0$, i.e. $\mu\propto v_0$, for which one recovers the formalism in this paper, as is clear from making $\mu \xi^{-\frac{D-2}{2}}=v_0=\phi_0^{reg}, \,\sigma=\tilde\phi^{reg}$ in eq.~\eqref{eq:muphi0}. Note that in the formalism here the equation expressing that the Green's functions have the correct mass dimension is not imposed somewhat ad-hoc, separately from the Ward identity, but rather coincides with the Ward Identity itself, as commented around eq.~\eqref{eq:Wardscaling}.

When discussing the physical relevance of the dilatation symmetry, it is instructive to point out the differences with the situation  in gravity theories in which  global scale invariance  is promoted to local Weyl invariance. The latter symmetry has to be gauge-fixed, and one can always choose a unitary gauge in which $\phi$ is a constant, so that a Weyl-invariant theory of scalar gravity is physically equivalent to a a theory of gravity with no dilaton and no Weyl symmetry \cite{Fradkin:1978yw,Tsamis:1984hh}. This result is valid for pure gravity theories as well as those including matter fields. Following \cite{Tsamis:1984hh}, a Weyl-invariant action involving the metric, gauge fields, fermions and scalars can always be written as
\begin{align}
\label{eq:S_equivalence}
 S[\phi,g,H,\psi,A]={\mathfrak S}\left[(\phi)^\frac{4}{D-2}g,\phi^{-1}H,\phi^{-\frac{D-1}{D-2}}\psi,A\right].
\end{align}
This allows to establish a physical equivalence between the theory defined by $S[\phi,g,H,\psi,A]$ and a theory with no dilaton and no Weyl symmetry with action ${\mathfrak S}\left[g,H,\psi,A\right]$, by choosing the gauge $\phi=1$. This equivalence is exact, valid at the quantum level, implying that Green's functions of Weyl- and diffeomorphism-invariant operators in the theory with action $S$ are identical to corresponding Green's functions of diffeomorphism-invariant operators in the theory with action $\mathfrak S[g,H,\phi,A]$. This  implies in particular that 
the existence of a nontrivial scaling of interaction strengths with energy, being a physical effect, should appear in both types of theories, and as already noted in refs.~\cite{Tsamis:1984hh,Antoniadis:1984br}, in the scale-invariant theory the renormalization scale $\mu$, along with all the other scales present in the dilatonless theory, should be traded for $\phi_0$. 

Given this, the result about having a nontrivial running despite the absence of a scale anomaly is not a surprise, though it does not follow from an equivalence between theories with and without a dilaton field: the above correspondence only applies to scale-invariant theories embeddable into Weyl-invariant ones, and also on the dilaton side it only applies to Green's functions of Weyl-invariant operators. Therefore, though in gravity theories the Weyl symmetry can be called, using the wording of ref.~\cite{Tsamis:1984hh}, a sham symmetry, the dilatation symmetry in flat-space theories  still is of physical consequence. As discussed above, the higher order interactions generated by the expansion of the evanescent powers of the dilaton will be of physical relevance, distinguishing these theories from those defined with an anomalous regularization, as happens already at the level of the effective potential. However, several issues should be pointed out. First, these scale-invariant theories do not seem to be ultraviolet-complete due to their nonrenormalizability. Furthermore, it is not clear how to deal with the unbroken phase, if indeed it is possible; for example, perturbation theory is only feasible in the broken phase, in which the evanescent powers of the dilaton can be expanded around its VEV. Finally, in scenarios in which the dilaton interacts weakly and has a very large VEV (see next section) the deviations between the predictions of these scale-invariant theories and those obtained using the traditional regularization, being suppressed by inverse powers of the VEV or by powers of the dilaton couplings, might be unaccessible to experiments if the VEV is large enough.


\section{Tuning and naturalness\label{sec:tuning}}

Scale-invariant theories such as those analyzed here have been hailed as a possible solution to the Higgs naturalness problem because the Higgs mass should be protected by the dilatation symmetry \cite{Shaposhnikov:2008xi}. Similarly, it has been advocated that these scale-invariant constructions can also solve the cosmological constant problem, since scale or Weyl symmetry forces the potential to have a minimum at zero \cite{Wetterich:1987fm,Shaposhnikov:2008xi}.\footnote{This can be easily seen from the fact that conformal symmetry and the absence of external scales forces the potential $V(\phi,H)$ to have mass dimension $D$, $DV=\phi\frac{\partial V}{\partial \phi}+H\frac{\partial V}{\partial H}$ so that at the minimum $\frac{\partial V}{\partial \phi}=\frac{\partial V}{\partial H}=V=0$.}  

The previous arguments, though, do not exclude the need to fine-tune dimensionless parameters. First, one may generalize the arguments given around  eq.~\eqref{eq:Vdilaton} to show that there is an implicit tuning in the choice of a potential with a flat direction, such as in eq.~\eqref{eq:Sdilaton}. Scale invariance, or Weyl invariance in the theory incorporating gravity, forces the potential to take the form \cite{Wetterich:1987fm,Shaposhnikov:2008xi}
\begin{align*}
 V(H,\phi)=\phi^{\frac{2D}{D-2}}F(H/\phi).
\end{align*}
If the vacuum is to have $\phi\neq0$, as needed to make the scale-invariant regularization well defined, then one needs a flat direction in the potential: $F(H/\phi)=0=F'(H/\phi)$ in the vacuum. This implies a tuning of parameters, coming from ensuring that $F$ has a critical point with value zero. To see this explicitly, note that the potential of eq.~\eqref{eq:Sdilaton} can be written in the form
\begin{align}
\label{eq:Vgeneral}
 V=\frac{\lambda}{4!}H^4+\frac{\lambda_\phi}{4!}\phi^4+\frac{\lambda_{H\phi}}{4}H^2\phi^2.
\end{align}
In terms of the couplings in the previous equation, the choice $\eta=0$ ensuring a flat direction in the potential of eq.~\eqref{eq:Sdilaton} is equivalent to 
\begin{align}
\label{eq:flat_d_condition}
 \lambda_{H\phi}<0,\quad \lambda_{H\phi}^2=\frac{1}{9}\lambda\lambda_\phi \,\,\Rightarrow \lambda_\phi=\lambda \zeta^4,\,\,\lambda_{H\phi}=-\frac{1}{3}\lambda\zeta^2,
\end{align}
which allows to write the tree-level potential in terms of just two couplings, $\lambda$ and $\zeta$. This condition, guaranteeing the existence of a flat direction at tree level, is not protected by any symmetry and is therefore not generic, i.e., it implies a tuning. Quantum corrections in principle modify the requirement for a flat direction, but its existence can always be guaranteed with proper renormalization conditions. Indeed, the three parameters $\lambda,\,\lambda_\phi,\,\lambda_{H\phi}$ are not predictions of the theory, and in the process of renormalization they have an associated ambiguity in the subtraction of finite parts, which is fixed by imposing three renormalization conditions. A flat direction can be guaranteed with one of these conditions; this was done explicitly at one loop in ref.~\cite{Shaposhnikov:2008xi}, and there is no obstruction to repeat the procedure at all orders of perturbation theory. Thus, the potential guaranteeing a flat direction is intrinsically tuned by imposing appropriate renormalization conditions, which at tree level are of the form \eqref{eq:flat_d_condition} \footnote{Note that this is a true tuning of dimensionless parameters rather than a relation allowing to trade dimensionless parameters for VEVs, as in dimensional transmutation (see the discussion around eq.~\eqref{eq:transmutation}). In contrast to the latter case, once the parameters are adjusted there is a flat direction and hence for a fixed choice of couplings the VEVs are not uniquely defined.}. This tuning associated with the existence of a vacuum with nonzero values for the dilaton can be interpreted as nothing but the cosmological constant problem rewritten in new variables, as noted in ref.~\cite{Tsamis:1984hh} in the context of pure scalar gravity with Weyl invariance. Indeed, given in that case the physical equivalence between a Weyl-invariant scalar-gravity theory coupled to matter and an ordinary theory of gravity and matter with no dilaton, the fact that the latter suffers from a cosmological constant problem forcing some heavy tuning implies that similar troubles must plague the former theory, albeit showing up in terms of dimensionless couplings. Though in the case without Weyl symmetry there are not such arguments of physical equivalence to theories without scale invariance, one has the same kind of undeniable fine tuning of dimensionless parameters as in Weyl-invariant theories.

What about the Higgs naturalness problem in the flat-space theory? Again, in the absence of Weyl symmetry there is in principle no physical equivalence with dilatonless theories. However, the  naturalness problem  is not necessarily solved if it is understood as the requirement of no unnatural constraints or cancellations in models which include generic heavy states. To show that this is not necessarily the case, let's consider eq.~\eqref{eq:Sdilaton} as a toy model for the potential in a theory in which $H$ can be thought of as the Higgs field. As was seen, acceptable phenomenology requires a very small coupling $\zeta$, affecting the dilaton-dilaton and dilaton-Higgs interactions, in order to recover the usual running. In the notation of eq.~\eqref{eq:Vgeneral}, this implies $\lambda^2\gg\lambda^2_{H\phi}\gg\lambda^2_\phi$. Since the square of the mass of the Higgs is proportional to $\lambda\zeta^2 \phi_0^2$, in order to achieve the correct value while keeping $\lambda$ perturbative one needs a large value of $\phi_0$, which could be taken to be of the order of some fundamental scale such as the Planck scale. In this case one would have, for $\lambda$ of order one, $\zeta\sim10^{-16}$, or $\lambda_{H\phi}\sim10^{-32},\lambda_{\phi}\sim10^{-64},$ and one may ask whether the smallness of these couplings can be naturally maintained in the presence of heavy states. For this, consider a heavy field $F$ which couples to the Higgs and the dilaton in a scale-invariant way (assuming again appropriate discrete symmetries):
\begin{align*}
V=\lambda_F\frac{1}{4!}F^4+\frac{1}{4}\lambda_{F\phi}F^2\phi^2+\frac{1}{4}\lambda_{F H}F^2 H^2+\frac{\bar\lambda}{4!}H^4+\frac{\bar\lambda_\phi}{4!}\phi^4+\frac{\bar\lambda_{H\phi}}{4}H^2\phi^2.
\end{align*}
The theory can still be regularized in a way preserving scale invariance, using analogues of the relation \eqref{eq:lambda0phi} for all the quartic couplings. $F$ can be made heavy by having a sizable, $O(1)$ coupling $\lambda_{F\phi}$; similarly, in a generic model one expects $\lambda_{F},\lambda_{FH}$ to be also of order one. Integrating F out using the equations of motion, the resulting low energy effective potential is again scale invariant and of the form of eq.~\eqref{eq:Vgeneral}, with the couplings given by
\begin{align*}
 \lambda&=\bar\lambda+9\frac{\lambda_{FH}^2}{\lambda_F},\\
 \lambda_\phi&=\bar\lambda_\phi+9\frac{\lambda_{F\phi}^2}{\lambda_F},\\
 \lambda_{H\phi}&=\bar\lambda_{H\phi}+3\frac{\lambda_{FH}\lambda_{F\phi}}{\lambda_F}.
\end{align*}
As was seen before, successful phenomenology (small Higgs-dilaton couplings and a flat direction) demands $\lambda^2\gg\lambda^2_{H\phi}\gg\lambda^2_\phi$. In particular, for $\phi_0$ of the order of the Planck scale, one has to have
\begin{align}
\label{eq:tuninglambdas}
 \left|\bar\lambda_{H\phi}+3\frac{\lambda_{FH}\lambda_{F\phi}}{\lambda_F}\right|\sim O(10^{-32}), \quad\bar\lambda_\phi+9\frac{\lambda_{F\phi}^2}{\lambda_F}\sim O(10^{-64}),
\end{align}
which, for order one values of the quartic couplings of the heavy field $F$, can only be achieved with substantial tuning. Note that this tuning is not even technically natural (protected by an enhanced symmetry when the couplings tend to zero) since eq.~\eqref{eq:tuninglambdas} implies that $\lambda_{H\phi}, \lambda_\phi$ in the UV theory have to be of the order of $\lambda_{FH},\lambda_{F\phi},\lambda_{F}$, which are $O(1)$ themselves, and yet finely tuned with respect to them. Therefore a scale-invariant construction is not necessarily natural; naturalness requires the hypothesis of the absence of generic new Physics coupling to the Higgs.

\section{Summary and conclusions\label{sec:conclusions}}

It has been shown that running couplings, rather than being associated with a scale anomaly, can be understood as arising from the spontaneous breaking of an anomaly-free dilatation symmetry in an appropriate effective theory involving a dilaton scalar with a nonzero VEV, which generates all mass scales. 

These theories are defined using the scale-symmetry-preserving regularization of ref.~\cite{Englert:1976ep}, and the conclusions were reached by considering scaling Ward identities, analogues of the Callan-Symanzik equations, and using the quantum action principle in dimensional regularization. No explicit renormalization scale was introduced at any point, so that dilatation invariance was preserved at all stages of the derivation. 

As it turns out, any ordinary theory seems to be embeddable into a scale-invariant theory by coupling it to a dilaton, as long as the latter has an exact flat direction. If the dilaton is sufficiently weakly coupled, albeit still generating sizable mass scales, one recovers the same running with energy of the effective couplings as in the theory without the dilaton but with massive parameters.

The issue of compatibility between running and exact dilatation symmetry had been previously tackled in the more restricted context of Weyl-invariant theories in refs.~\cite{Antoniadis:1984kd,Antoniadis:1984br}; however, in contrast to the treatment here, the formalism in those references involved the introduction of an explicit renormalization scale, which in principle generates an anomaly which has to be eliminated by imposing additional conditions, which were not correctly formulated. Similarly, ref.~\cite{Codello:2012sn} studied the renormalization group flow in classically Weyl-symmetric theories which were regularized using dimensionful cutoffs proportional to the dilaton VEV; however, given that physical predictions should be cutoff-independent, such a regularization is questionable (this would be similar to choosing $\mu=\zeta\phi_0$ in DR, which does not get rid of the anomaly).

Arguments supporting  the recovery of ordinary effective running couplings in the scale-invariant theories studied here had been given in ref.~\cite{Shaposhnikov:2008xi} and were supported by a one-loop computation of the four-point Green's function in the theory of a scalar field coupled to the dilaton; however, it was unclear whether the result would hold at higher orders or for arbitrary Green's functions, and whether one could define beta functions allowing to resum the corrections with a logarithmic dependence in momenta. These things were shown to be true in this paper for theories with scalars, fermions and gauge fields.

The results presented here contradict the commonly accepted knowledge that running couplings can only appear as a consequence of a scale anomaly, showing that they may arise from the spontaneous breaking of an exact scale invariance. 
Nevertheless, this is not entirely surprising given known results for gravitational theories that establish the physical equivalence of scale-free Weyl-symmetric theories of scalar gravity with dilaton-free theories of gravity \cite{Fradkin:1978yw,Tsamis:1984hh}, implying that the running couplings of the latter must somehow make an appearance in the former. How such a thing may take place was clarified even for flat-space theories for which the former equivalence does not apply; ultimately, the nontrivial scaling is due to the evanescent (vanishing in $D=4$ dimensions) power of the dilaton field in the interactions, so that the usual renormalization scale $\mu$ is traded for a scale proportional to the dilaton VEV. 

Some aspects of these scale-invariant theories deserve further study. As has been emphasized, they seem to make sense only as effective theories, being nonrenormalizable \cite{Shaposhnikov:2009nk}, and it is unclear whether the unbroken phase is well defined. It would be interesting to understand better the physical consequences of the exact scale invariance. As was noted, there are deviations from the behavior of classically scale-invariant theories regularized in the usual way; although some of the differences in physical predictions could be negligible for phenomenologically interesting scenarios with very large dilaton VEVs, this is model-dependent, and in any case there can be nontrivial consequences regarding vacuum stability --for example, the effective potential is restricted by exact scale invariance, which was seen to provide protection from destabilizing effects by fermions. Finally, although these theories realize a dynamical mechanism for the generation of mass scales through the spontaneous breaking of scale invariance, they are not free of tuning problems. First, the need for a flat direction implies a tuning which can be interpreted along the lines of ref.~\cite{Tsamis:1984hh} as a rephrasing of the cosmological constant problem in terms of dimensionless couplings. Lastly, in the presence of heavy states there is still a Higgs naturalness problem. 

To conclude, it should be commented that the results obtained here should not be viewed as an  artifact of dimensional regularization. Similar conclusions should be reached by using other regularization schemes, although the derivations would become less transparent. In particular, it should be possible to construct effective theories with spontaneously broken exact scale invariance using different regularization methods, modifying the Lagrangian with the insertion of additional, nonlocal, cutoff-dependent dilaton interactions that cancel the anomaly and become trivial in the limit in which the regularization is removed, and which in the broken phase would generate an infinite number of local terms, playing the same role as the terms coming from expanding the evanescent powers of the dilaton in DR. 
\newpage
\section*{Acknowledgements}
I thank Itay Yavin for turning my attention into these theories and for extended discussions; thanks also to Brian Shuve and Benjam\'{\i}n Grinstein for useful exchanges. I acknowledge support from the Spanish Government through grant FPA2011-24568. Research at the Perimeter Institute is supported in part by the Government of Canada through Industry Canada, and by the Province of Ontario through the Ministry of Research and Information (MRI).

\bibliographystyle{h-physrev}
\bibliography{bib-Scale_invariant_theories}

\end{document}